\documentclass[aps,prl,twocolumn,groupedaddress,amsmath,amssymb]{revtex4-1}
\usepackage{graphicx}  
\usepackage{dcolumn}   
\usepackage{bm}        
\usepackage{verbatim}   

\begin{document}
\title{Crystallization and glass transition in supercooled binary Lennard-Jones liquids}

\author{Penghui Cao}
  \email[]{pcao@mit.edu}
  \affiliation{Department of Nuclear Science and Engineering, Massachusetts Institute of Technology,Cambridge, Massachusetts 02139}

\date{\today}

\begin{abstract}

The classic Kob-Andersen (KA) binary Lennard-Jones mixtures which are designed to prevent crystallization has been extensively studied in simulation of slow dynamics. Although crystallization can occur if a liquid system is cooled slowly, so far the  KA model has not been crystallized. Here we report using molecular simulation the observation of crystal growth in the supercooled KA liquids. The onset of crystallization is observed occurring at temperature $T_c= 0.55$ which is higher than the  glass transition temperature of $T_g=0.40$. We further examine the statistical distribution of single particle displacements in crystallization and close to glass transition. The displacement distribution for crystallization exhibits a power-law decay, whereas the distribution for glassy relaxation reflects a Gaussian center, terminated with an exponential tail (namely dynamic heterogeneity). Finally, we predict in order to crystalize KA liquids the cooling rate is approximately equal to $10^{-22}$, which is about 15 odder lower than the typical MD cooling rate. 

\end{abstract}

\maketitle

Glasses  are typically prepared by quick cooling from their liquid states. The liquid viscosity rapidly increases by many orders of magnitude when approaching glass transition, and significantly slows down the atom motion. The atoms rearrange so slow that they do not have sufficient time to sample many configurations within a short time window. Therefore the quickly quenched configuration appears as disordered, but behaves mechanically like solids. The classic KA mixture \cite{kob1995testing,kob1995testing2}  that emerged as model to probe glass transition has recently been extensively used to study slow dynamics in supercooled liquid and mechanical properties of glasses \cite{sastry1998signatures,donati1998stringlike,angelani2000saddles,shi2006atomic,kushimaJCP2009,rodney2011modeling,caoPRE2012,singh2013ultrastable, caoJMPS2014}. The parameters in the Lennar-Jones potential were adjusted, making the model less prone to crystallization. At a sufficiently low cooling rate, one may expect the mixtures could still form crystalline structures. There has been a number of studies by different groups, aimed to understand crystallization of KA model \cite{doye2007controlling,toxvaerd2009stability,banerjee2013interplay}. However, crystallization of this mixture are reported only when changing their composition or inter-species interaction length. To the best of the authors' knowledge, the crystallization behavior of the original KA model remains unknown. 

\begin{figure} \begin{center}
\includegraphics[scale=0.525]{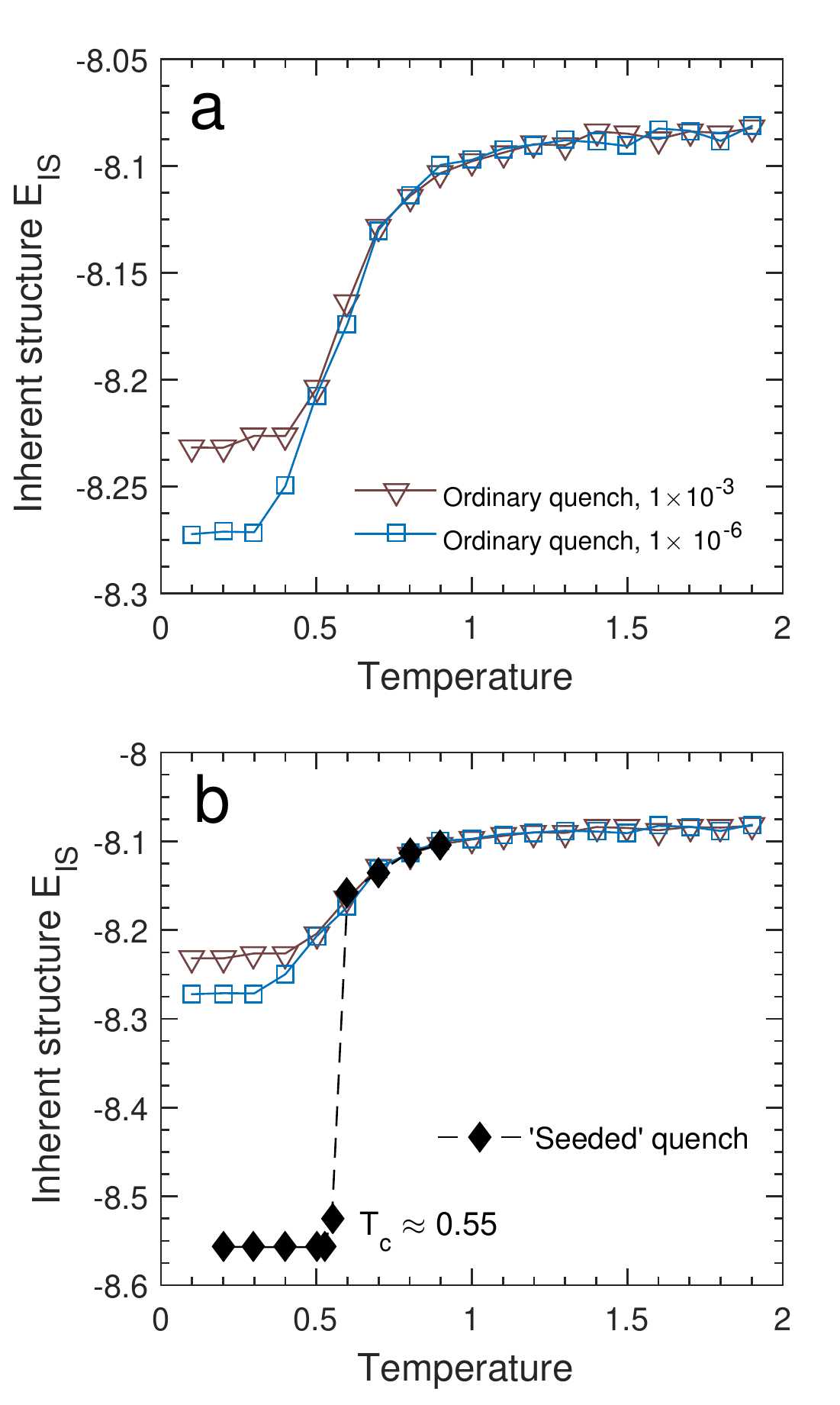} 
\caption{(Color online) The averaged inherent structure energy per particle $E_{is}$ as a function temperature T. (a) Ordinary MD quenching at cooling rates of $1\times10^{-1}$ and $1\times10^{-6}$. Each data points represent an average of 50 inherent structures. (b) Comparison between ordinary quench and seeded quench.  The significant energy drop at $T=0.55$  indicates phase transition from liquid to crystal. }
\label{fig1} \end{center} \end{figure}

\begin{figure*} \begin{center}
\includegraphics[scale=0.38]{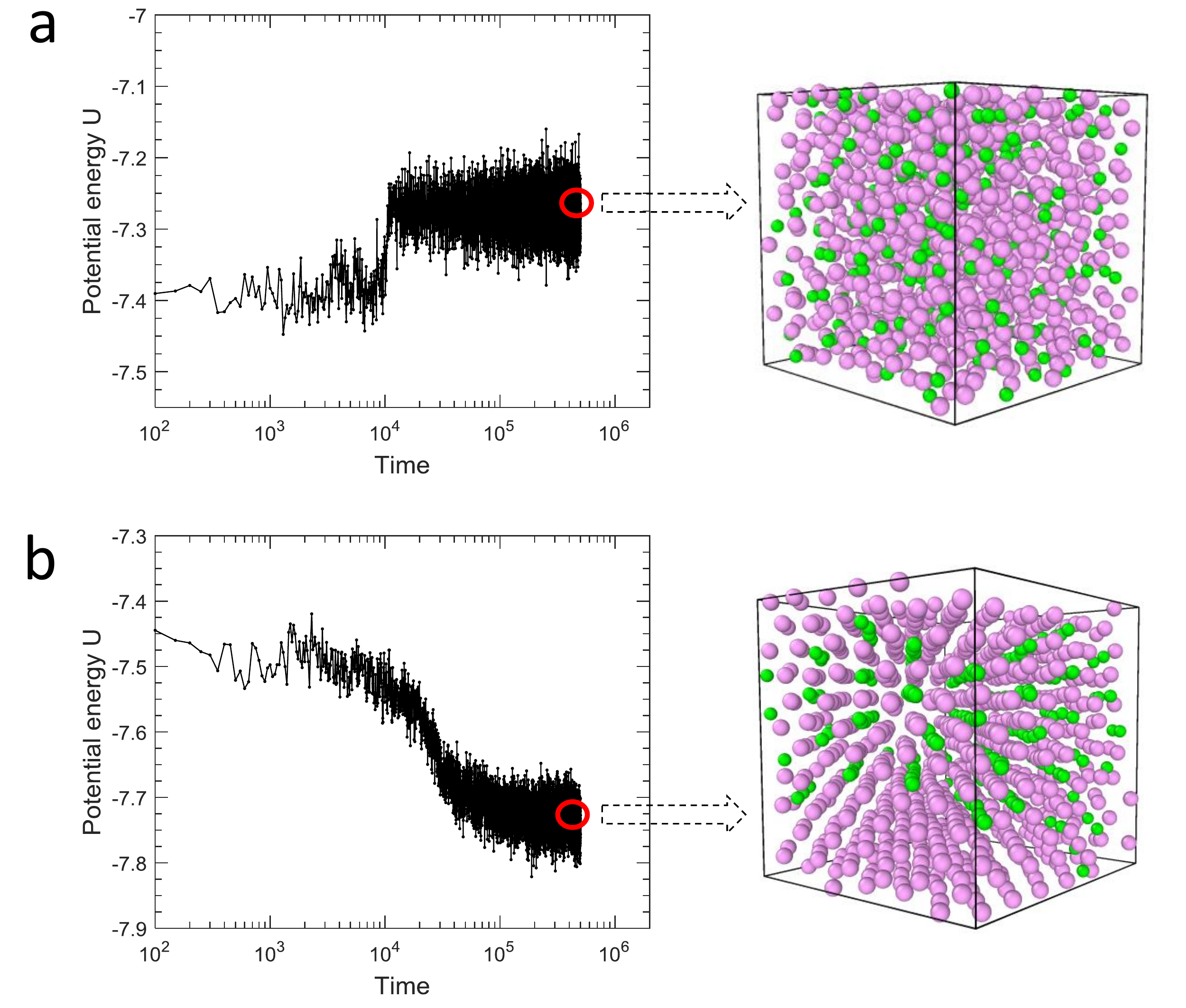} 
\caption{(Color online) (a) Time evolution of potential energy during seeded quenching at temperatures (a) $T = 0.6$ and (b) $T = 0.55$. The process of crystallization is observed at $T=0.55$.}
\label{fig2} \end{center} \end{figure*}

To this end, we performed molecular dynamics (MD) simulations of glass transition and crystal growth in the supercooled KA liquids. The system contains 1,000 particles with the same unit mass; 80$\%$ of the particles are of type A, 20 are type B. The particles interacts through the binary Lennard-Jones potential and the parameters can be found in this original paper \cite{kob1995testing}. The atomic interaction is truncated at a cutoff distance of $r_{\alpha\beta} = 2.5\sigma_{\alpha\beta}$, where $\alpha, \beta$ equal to A or B. Periodic boundary conditions are applied in all three directions and the particle density is fixed at $\rho = 1.2$. Here, we use two quenching protocols to simulate glass transition and crystal growth. For the simulation of glass transition, we follow the quenching protocol as in \cite{sastry1998signatures}. The system is initially equilibrated at high temperature $T = 2.0$ and then further quenched to a low temperature of 0.1 at a given cooling rate. The inherent structure energy $E_{IS}$ is computed by performing energy minimization of configurations quenched at each temperature. Atomistic modeling of crystallization for supercooled liquids is a challenge due to the time-scale limitation of the MD method. It's known that the rate of crystallization for a liquid depends on crystal nucleation rate and growth rate. Crystal nucleation process usually takes much longer time as comparing with growth \cite{tang2013anomalously}. To facilitate the crystallization modeling, a small crystal ``seed" is manually created and embedded in the liquid configuration during quenching. This proposed protocol is named as ``seeded"quenching which enables the investigation of crystal growth in supercooled liquids.

Fig.1 shows the temperature dependent inherent structure energy obtained using the above two quenching protocols, ordinary quenching and seeded quenching The ordinary cooling is performed with two cooling rates of  $1\times10^{-3}$  and $1\times10^{-6}$.  At the high temperature ($T > 1.0$) and low temperature ($T<0.3$) regions, the inherent structure energies do not change much with temperature. Between $T = 1.0$ and $T = 0.3-0.4$, those energies decrease dramatically. The glass configuration prepared at slower rate has a lower potential energy, which suggests slow cooling enables the system to access deep energy basins, and the  deeply trapped glasses are  comparatively stable. The glass transition temperature is estimated to be about 0.40 based on the shape of the energy curve. These results and observations in normal cooling simulations are in consistent with previous studies \cite{sastry1998signatures}. Fig. 1(b) presents the comparison the energy plots  between ordinary quenching and seeded quenching. As one can see from it, the energy curve of seeded quenching basically follows the same trend as the normal quenching at temperature $T>0.55$, that indicates liquid phase of the system and that the embedded seed has been dissolved due to high thermal energy. Interestingly, as decrease temperature $T = 0.55$, we observe a significant energy drop, implying a phase transition from liquid state to crystal state. This crystallization transition is further confirmed by examining the atomic structure as shown in Fig. 2(b).  This crystal configuration has a potential energy of  -8.56 at the density of 1.2 that is significantly lower than the reported ultrastable glass in previous study \cite{singh2013ultrastable}.

\begin{figure} \begin{center}
\includegraphics[scale=0.48]{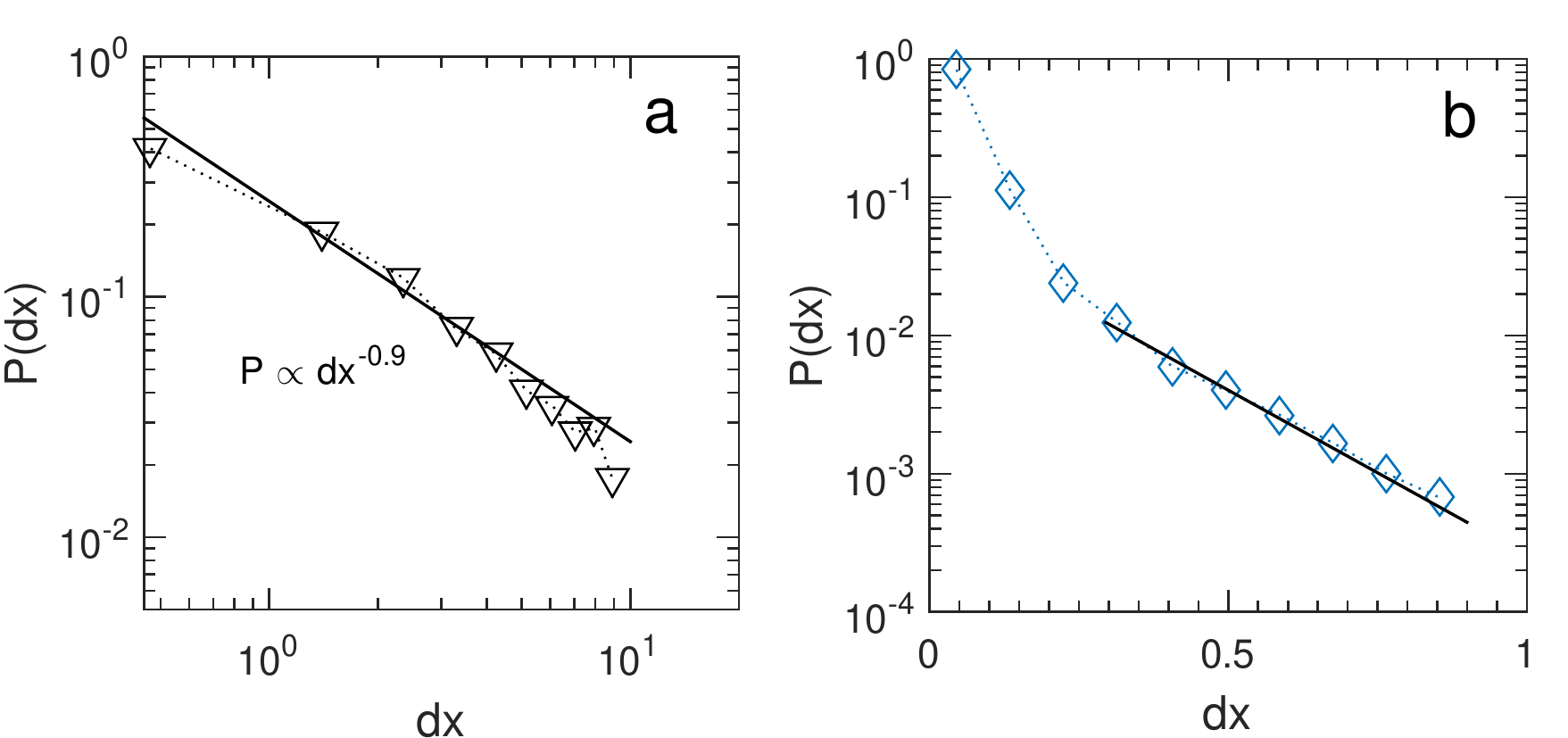} 
\caption{(Color online) Probability distributions of particle displacements in crystallization and close to glass transition. (a) Crystallization exhibits a power-law distribution with exponent -0.9 at $T = 0.55$. (b) the relaxation of supercooled liquids at $T = 0.4$  shows Gaussian center with an exponential decaying. } 
\end{center} \end{figure}

Fig. 2 presents time evolution of system potential energy during seeded quenching at temperatures of $T=0.6$ and $T=0.5$. We briefly describe the seeded quenching method: a small FCC crystal containing 40 A atoms is implanted into the supercooled liquids. We relax the system for t = 500 with the seed pinned, and then set free the pinned atoms and relax the whole system for another $5\times10^5$. In Fig. 2(a), one can see there is a potential energy jump taking place at time of t = $10^4$ at temperature $T=0.6$, that is ascribed to dissolution of the crystal seed. The final configuration showing disordered structure maintains its liquid state. However, the system potential energy progressively decreases at the temperature $T = 0.55$ as shown in Fig. 2(b). We find the energy drop at time $t = 8\times10^3$ corresponds to the growth of crystal seed.  The crystallization of the entire system is accomplished at time $t = 10^5$ and the  finalconfiguration shows a mixed crystal.

\begin{figure} \begin{center}
\includegraphics[scale=0.3]{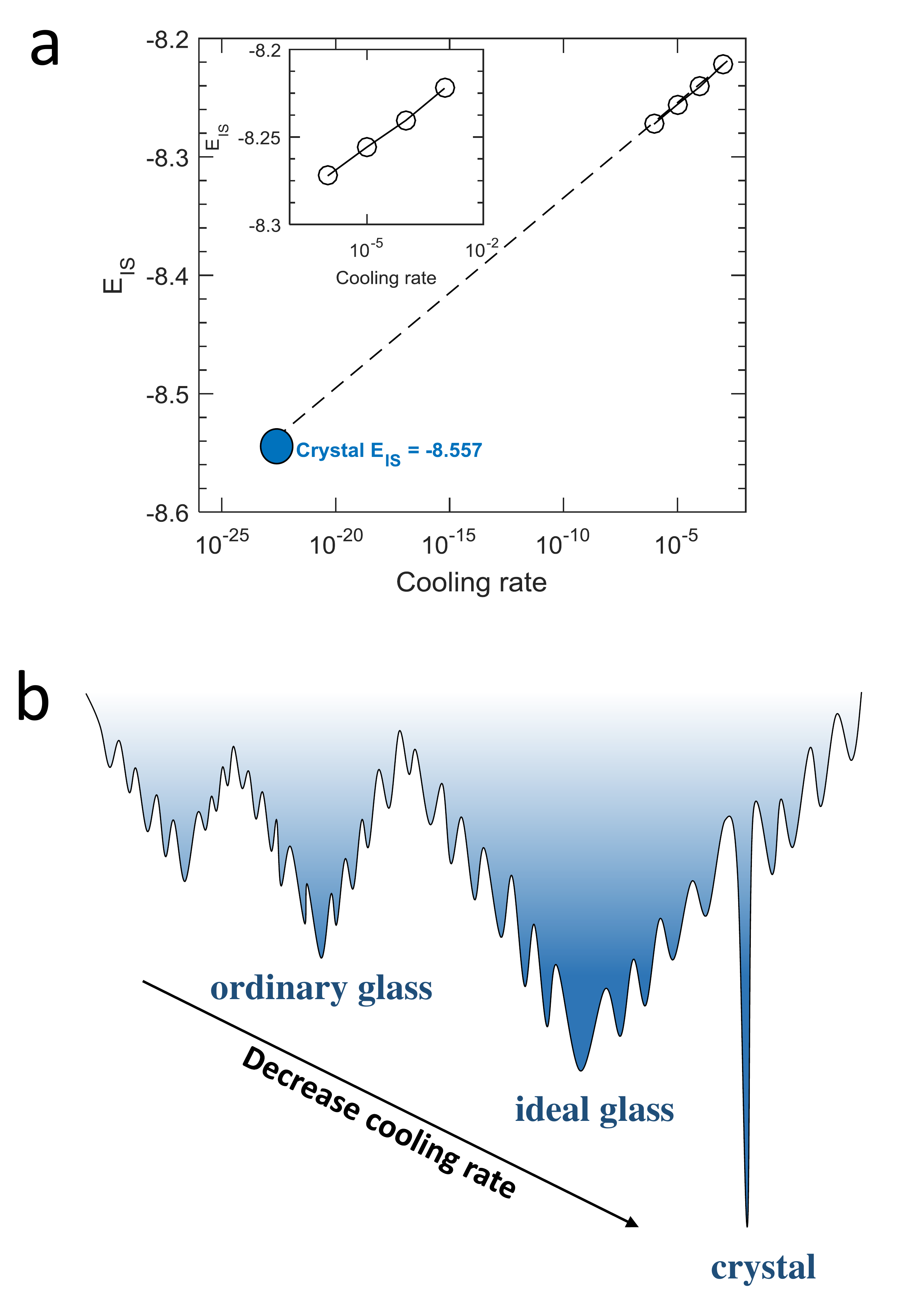} 
\caption{(Color online) (a) Cooling rate dependence of inherent structure energy and predicted crystallization cooling rate of $1\times10^{-22}$.  The inset highlights MD results of a linear decrease of inherent energy with cooling rate. (b) A schematic illustration of potential energy landscape. The horizontal axis represents configuration coordinates and vertical axis marks potential energy.}
\end{center} \end{figure}

To characterize relaxation dynamics near glass transition and in crystallization, we have investigated the distribution of single particle displacements, shown in Fig. 3. The relaxation of glassy materials close to glass transition temperature exhibits a Gaussian core, which corresponds to localized slow particle motion, and an exponential tail responding to the high mobility parties. The coexistence of the slow and fast particles reflects the dynamical heterogeneity in glassy material relaxation \cite{Kob2007universal,weeks2000three}.  In contrary to glass relaxation, crystallization presents a nearly power-law distribution in particle displacement. We suggest that the crystallization process can be viewed as a critical phenomenon \cite{bak1988self,turcotte1999self,sethna2001crackling}.

Finally, the cooling rate required for crystallization is estimated by a simple extrapolation. As seen in the inset of Fig. 4 (a), the inherent energy of prepared glass linearly decreases with cooling rate. Following this trend, we predict the crystallization cooling rate is approximately $10^{-22}$, which is well beyond the capability of traditional MD simulation. We show in Fig. 4(b) a schematic plot of a potential energy landscape, illustrating how the quenched configurations depend on cooling history. As decrease cooling rate, one would expect the glasses become much more stable, though the energy is still higher than the magnitude of its crystal state. Deep into the energy well, the landscape is rough and includes high energy barriers separating the neighboring local energy minima. We consider these deeply trapped glass states relatively stable. As the cooling rate continues to decrease, the system gets enough time to sample many energy basins, and crystallization occurs at a rapid speed when flowing into the global energy minimum.

In summary, a seeded quenching protocol is introduced to simulate crystallization of supercooled liquids. We found in this work the suppercooled KA mixtures are able to crystallize at temperature of 0.55. The statistical distribution of single particle displacement during the process of crystallization exhibits power-law decay, implying the self-organized criticality behavior. The distribution of displacement in glassy relaxation approaching to glass transition shows the coexistence of slow motion and fast travel particles, indicating the dynamical heterogeneity in glassy slow dynamics. We found the inherent energy of prepared glasses linearly decreases with cooling rate. Using simple energy extrapolation, we estimate the cooling rate to suffice crystallization of KA mixture to be $10^{-22}$.      

\bibliography{biball}

\begin{thebibliography}{20}%
\makeatletter
\providecommand \@ifxundefined [1]{%
 \@ifx{#1\undefined}
}%
\providecommand \@ifnum [1]{%
 \ifnum #1\expandafter \@firstoftwo
 \else \expandafter \@secondoftwo
 \fi
}%
\providecommand \@ifx [1]{%
 \ifx #1\expandafter \@firstoftwo
 \else \expandafter \@secondoftwo
 \fi
}%
\providecommand \natexlab [1]{#1}%
\providecommand \enquote  [1]{``#1''}%
\providecommand \bibnamefont  [1]{#1}%
\providecommand \bibfnamefont [1]{#1}%
\providecommand \citenamefont [1]{#1}%
\providecommand \href@noop [0]{\@secondoftwo}%
\providecommand \href [0]{\begingroup \@sanitize@url \@href}%
\providecommand \@href[1]{\@@startlink{#1}\@@href}%
\providecommand \@@href[1]{\endgroup#1\@@endlink}%
\providecommand \@sanitize@url [0]{\catcode `\\12\catcode `\$12\catcode
  `\&12\catcode `\#12\catcode `\^12\catcode `\_12\catcode `\%12\relax}%
\providecommand \@@startlink[1]{}%
\providecommand \@@endlink[0]{}%
\providecommand \url  [0]{\begingroup\@sanitize@url \@url }%
\providecommand \@url [1]{\endgroup\@href {#1}{\urlprefix }}%
\providecommand \urlprefix  [0]{URL }%
\providecommand \Eprint [0]{\href }%
\providecommand \doibase [0]{http://dx.doi.org/}%
\providecommand \selectlanguage [0]{\@gobble}%
\providecommand \bibinfo  [0]{\@secondoftwo}%
\providecommand \bibfield  [0]{\@secondoftwo}%
\providecommand \translation [1]{[#1]}%
\providecommand \BibitemOpen [0]{}%
\providecommand \bibitemStop [0]{}%
\providecommand \bibitemNoStop [0]{.\EOS\space}%
\providecommand \EOS [0]{\spacefactor3000\relax}%
\providecommand \BibitemShut  [1]{\csname bibitem#1\endcsname}%
\let\auto@bib@innerbib\@empty
\bibitem [{\citenamefont {Kob}\ and\ \citenamefont
  {Andersen}(1995{\natexlab{a}})}]{kob1995testing}%
  \BibitemOpen
  \bibfield  {author} {\bibinfo {author} {\bibfnamefont {W.}~\bibnamefont
  {Kob}}\ and\ \bibinfo {author} {\bibfnamefont {H.~C.}\ \bibnamefont
  {Andersen}},\ }\href@noop {} {\bibfield  {journal} {\bibinfo  {journal}
  {Physical Review E}\ }\textbf {\bibinfo {volume} {51}},\ \bibinfo {pages}
  {4626} (\bibinfo {year} {1995}{\natexlab{a}})}\BibitemShut {NoStop}%
\bibitem [{\citenamefont {Kob}\ and\ \citenamefont
  {Andersen}(1995{\natexlab{b}})}]{kob1995testing2}%
  \BibitemOpen
  \bibfield  {author} {\bibinfo {author} {\bibfnamefont {W.}~\bibnamefont
  {Kob}}\ and\ \bibinfo {author} {\bibfnamefont {H.~C.}\ \bibnamefont
  {Andersen}},\ }\href@noop {} {\bibfield  {journal} {\bibinfo  {journal}
  {Physical Review E}\ }\textbf {\bibinfo {volume} {52}},\ \bibinfo {pages}
  {4134} (\bibinfo {year} {1995}{\natexlab{b}})}\BibitemShut {NoStop}%
\bibitem [{\citenamefont {Sastry}\ \emph {et~al.}(1998)\citenamefont {Sastry},
  \citenamefont {Debenedetti},\ and\ \citenamefont
  {Stillinger}}]{sastry1998signatures}%
  \BibitemOpen
  \bibfield  {author} {\bibinfo {author} {\bibfnamefont {S.}~\bibnamefont
  {Sastry}}, \bibinfo {author} {\bibfnamefont {P.~G.}\ \bibnamefont
  {Debenedetti}}, \ and\ \bibinfo {author} {\bibfnamefont {F.~H.}\ \bibnamefont
  {Stillinger}},\ }\href@noop {} {\bibfield  {journal} {\bibinfo  {journal}
  {Nature}\ }\textbf {\bibinfo {volume} {393}},\ \bibinfo {pages} {554}
  (\bibinfo {year} {1998})}\BibitemShut {NoStop}%
\bibitem [{\citenamefont {Donati}\ \emph {et~al.}(1998)\citenamefont {Donati},
  \citenamefont {Douglas}, \citenamefont {Kob}, \citenamefont {Plimpton},
  \citenamefont {Poole},\ and\ \citenamefont {Glotzer}}]{donati1998stringlike}%
  \BibitemOpen
  \bibfield  {author} {\bibinfo {author} {\bibfnamefont {C.}~\bibnamefont
  {Donati}}, \bibinfo {author} {\bibfnamefont {J.~F.}\ \bibnamefont {Douglas}},
  \bibinfo {author} {\bibfnamefont {W.}~\bibnamefont {Kob}}, \bibinfo {author}
  {\bibfnamefont {S.~J.}\ \bibnamefont {Plimpton}}, \bibinfo {author}
  {\bibfnamefont {P.~H.}\ \bibnamefont {Poole}}, \ and\ \bibinfo {author}
  {\bibfnamefont {S.~C.}\ \bibnamefont {Glotzer}},\ }\href@noop {} {\bibfield
  {journal} {\bibinfo  {journal} {Physical review letters}\ }\textbf {\bibinfo
  {volume} {80}},\ \bibinfo {pages} {2338} (\bibinfo {year}
  {1998})}\BibitemShut {NoStop}%
\bibitem [{\citenamefont {Angelani}\ \emph {et~al.}(2000)\citenamefont
  {Angelani}, \citenamefont {Di~Leonardo}, \citenamefont {Ruocco},
  \citenamefont {Scala},\ and\ \citenamefont
  {Sciortino}}]{angelani2000saddles}%
  \BibitemOpen
  \bibfield  {author} {\bibinfo {author} {\bibfnamefont {L.}~\bibnamefont
  {Angelani}}, \bibinfo {author} {\bibfnamefont {R.}~\bibnamefont
  {Di~Leonardo}}, \bibinfo {author} {\bibfnamefont {G.}~\bibnamefont {Ruocco}},
  \bibinfo {author} {\bibfnamefont {A.}~\bibnamefont {Scala}}, \ and\ \bibinfo
  {author} {\bibfnamefont {F.}~\bibnamefont {Sciortino}},\ }\href@noop {}
  {\bibfield  {journal} {\bibinfo  {journal} {Physical review letters}\
  }\textbf {\bibinfo {volume} {85}},\ \bibinfo {pages} {5356} (\bibinfo {year}
  {2000})}\BibitemShut {NoStop}%
\bibitem [{\citenamefont {Shi}\ and\ \citenamefont
  {Falk}(2006)}]{shi2006atomic}%
  \BibitemOpen
  \bibfield  {author} {\bibinfo {author} {\bibfnamefont {Y.}~\bibnamefont
  {Shi}}\ and\ \bibinfo {author} {\bibfnamefont {M.~L.}\ \bibnamefont {Falk}},\
  }\href@noop {} {\bibfield  {journal} {\bibinfo  {journal} {Physical Review
  B}\ }\textbf {\bibinfo {volume} {73}},\ \bibinfo {pages} {214201} (\bibinfo
  {year} {2006})}\BibitemShut {NoStop}%
\bibitem [{\citenamefont {Kushima}\ \emph {et~al.}(2009)\citenamefont
  {Kushima}, \citenamefont {Lin}, \citenamefont {Li}, \citenamefont {Eapen},
  \citenamefont {Mauro}, \citenamefont {Qian}, \citenamefont {Diep},\ and\
  \citenamefont {Yip}}]{kushimaJCP2009}%
  \BibitemOpen
  \bibfield  {author} {\bibinfo {author} {\bibfnamefont {A.}~\bibnamefont
  {Kushima}}, \bibinfo {author} {\bibfnamefont {X.}~\bibnamefont {Lin}},
  \bibinfo {author} {\bibfnamefont {J.}~\bibnamefont {Li}}, \bibinfo {author}
  {\bibfnamefont {J.}~\bibnamefont {Eapen}}, \bibinfo {author} {\bibfnamefont
  {J.~C.}\ \bibnamefont {Mauro}}, \bibinfo {author} {\bibfnamefont
  {X.}~\bibnamefont {Qian}}, \bibinfo {author} {\bibfnamefont {P.}~\bibnamefont
  {Diep}}, \ and\ \bibinfo {author} {\bibfnamefont {S.}~\bibnamefont {Yip}},\
  }\href@noop {} {\bibfield  {journal} {\bibinfo  {journal} {J. Chem. Phys.}\
  }\textbf {\bibinfo {volume} {130}},\ \bibinfo {pages} {224504} (\bibinfo
  {year} {2009})}\BibitemShut {NoStop}%
\bibitem [{\citenamefont {Rodney}\ \emph {et~al.}(2011)\citenamefont {Rodney},
  \citenamefont {Tanguy},\ and\ \citenamefont
  {Vandembroucq}}]{rodney2011modeling}%
  \BibitemOpen
  \bibfield  {author} {\bibinfo {author} {\bibfnamefont {D.}~\bibnamefont
  {Rodney}}, \bibinfo {author} {\bibfnamefont {A.}~\bibnamefont {Tanguy}}, \
  and\ \bibinfo {author} {\bibfnamefont {D.}~\bibnamefont {Vandembroucq}},\
  }\href@noop {} {\bibfield  {journal} {\bibinfo  {journal} {Modelling and
  Simulation in Materials Science and Engineering}\ }\textbf {\bibinfo {volume}
  {19}},\ \bibinfo {pages} {083001} (\bibinfo {year} {2011})}\BibitemShut
  {NoStop}%
\bibitem [{\citenamefont {Cao}\ \emph {et~al.}(2012)\citenamefont {Cao},
  \citenamefont {Li}, \citenamefont {Heugle}, \citenamefont {Park},\ and\
  \citenamefont {Lin}}]{caoPRE2012}%
  \BibitemOpen
  \bibfield  {author} {\bibinfo {author} {\bibfnamefont {P.}~\bibnamefont
  {Cao}}, \bibinfo {author} {\bibfnamefont {M.}~\bibnamefont {Li}}, \bibinfo
  {author} {\bibfnamefont {R.~J.}\ \bibnamefont {Heugle}}, \bibinfo {author}
  {\bibfnamefont {H.~S.}\ \bibnamefont {Park}}, \ and\ \bibinfo {author}
  {\bibfnamefont {X.}~\bibnamefont {Lin}},\ }\href@noop {} {\bibfield
  {journal} {\bibinfo  {journal} {Phys. Rev. E}\ }\textbf {\bibinfo {volume}
  {86}},\ \bibinfo {pages} {016710} (\bibinfo {year} {2012})}\BibitemShut
  {NoStop}%
\bibitem [{\citenamefont {Singh}\ \emph {et~al.}(2013)\citenamefont {Singh},
  \citenamefont {Ediger},\ and\ \citenamefont
  {De~Pablo}}]{singh2013ultrastable}%
  \BibitemOpen
  \bibfield  {author} {\bibinfo {author} {\bibfnamefont {S.}~\bibnamefont
  {Singh}}, \bibinfo {author} {\bibfnamefont {M.}~\bibnamefont {Ediger}}, \
  and\ \bibinfo {author} {\bibfnamefont {J.~J.}\ \bibnamefont {De~Pablo}},\
  }\href@noop {} {\bibfield  {journal} {\bibinfo  {journal} {Nature materials}\
  }\textbf {\bibinfo {volume} {12}},\ \bibinfo {pages} {139} (\bibinfo {year}
  {2013})}\BibitemShut {NoStop}%
\bibitem [{\citenamefont {Cao}\ \emph {et~al.}(2014)\citenamefont {Cao},
  \citenamefont {Lin},\ and\ \citenamefont {Park}}]{caoJMPS2014}%
  \BibitemOpen
  \bibfield  {author} {\bibinfo {author} {\bibfnamefont {P.}~\bibnamefont
  {Cao}}, \bibinfo {author} {\bibfnamefont {X.}~\bibnamefont {Lin}}, \ and\
  \bibinfo {author} {\bibfnamefont {H.~S.}\ \bibnamefont {Park}},\ }\href@noop
  {} {\bibfield  {journal} {\bibinfo  {journal} {J. Mech. Phys. Solids}\
  }\textbf {\bibinfo {volume} {68}},\ \bibinfo {pages} {239} (\bibinfo {year}
  {2014})}\BibitemShut {NoStop}%
\bibitem [{\citenamefont {Doye}\ \emph {et~al.}(2007)\citenamefont {Doye},
  \citenamefont {Louis}, \citenamefont {Lin}, \citenamefont {Allen},
  \citenamefont {Noya}, \citenamefont {Wilber}, \citenamefont {Kok},\ and\
  \citenamefont {Lyus}}]{doye2007controlling}%
  \BibitemOpen
  \bibfield  {author} {\bibinfo {author} {\bibfnamefont {J.~P.}\ \bibnamefont
  {Doye}}, \bibinfo {author} {\bibfnamefont {A.~A.}\ \bibnamefont {Louis}},
  \bibinfo {author} {\bibfnamefont {I.-C.}\ \bibnamefont {Lin}}, \bibinfo
  {author} {\bibfnamefont {L.~R.}\ \bibnamefont {Allen}}, \bibinfo {author}
  {\bibfnamefont {E.~G.}\ \bibnamefont {Noya}}, \bibinfo {author}
  {\bibfnamefont {A.~W.}\ \bibnamefont {Wilber}}, \bibinfo {author}
  {\bibfnamefont {H.~C.}\ \bibnamefont {Kok}}, \ and\ \bibinfo {author}
  {\bibfnamefont {R.}~\bibnamefont {Lyus}},\ }\href@noop {} {\bibfield
  {journal} {\bibinfo  {journal} {Physical Chemistry Chemical Physics}\
  }\textbf {\bibinfo {volume} {9}},\ \bibinfo {pages} {2197} (\bibinfo {year}
  {2007})}\BibitemShut {NoStop}%
\bibitem [{\citenamefont {Toxvaerd}\ \emph {et~al.}(2009)\citenamefont
  {Toxvaerd}, \citenamefont {Pedersen}, \citenamefont {Schr{\o}der},\ and\
  \citenamefont {Dyre}}]{toxvaerd2009stability}%
  \BibitemOpen
  \bibfield  {author} {\bibinfo {author} {\bibfnamefont {S.}~\bibnamefont
  {Toxvaerd}}, \bibinfo {author} {\bibfnamefont {U.~R.}\ \bibnamefont
  {Pedersen}}, \bibinfo {author} {\bibfnamefont {T.~B.}\ \bibnamefont
  {Schr{\o}der}}, \ and\ \bibinfo {author} {\bibfnamefont {J.~C.}\ \bibnamefont
  {Dyre}},\ }\href@noop {} {\bibfield  {journal} {\bibinfo  {journal} {The
  Journal of chemical physics}\ }\textbf {\bibinfo {volume} {130}},\ \bibinfo
  {pages} {224501} (\bibinfo {year} {2009})}\BibitemShut {NoStop}%
\bibitem [{\citenamefont {Banerjee}\ \emph {et~al.}(2013)\citenamefont
  {Banerjee}, \citenamefont {Chakrabarty},\ and\ \citenamefont
  {Bhattacharyya}}]{banerjee2013interplay}%
  \BibitemOpen
  \bibfield  {author} {\bibinfo {author} {\bibfnamefont {A.}~\bibnamefont
  {Banerjee}}, \bibinfo {author} {\bibfnamefont {S.}~\bibnamefont
  {Chakrabarty}}, \ and\ \bibinfo {author} {\bibfnamefont {S.~M.}\ \bibnamefont
  {Bhattacharyya}},\ }\href@noop {} {\bibfield  {journal} {\bibinfo  {journal}
  {The Journal of chemical physics}\ }\textbf {\bibinfo {volume} {139}},\
  \bibinfo {pages} {104501} (\bibinfo {year} {2013})}\BibitemShut {NoStop}%
\bibitem [{\citenamefont {Tang}\ and\ \citenamefont
  {Harrowell}(2013)}]{tang2013anomalously}%
  \BibitemOpen
  \bibfield  {author} {\bibinfo {author} {\bibfnamefont {C.}~\bibnamefont
  {Tang}}\ and\ \bibinfo {author} {\bibfnamefont {P.}~\bibnamefont
  {Harrowell}},\ }\href@noop {} {\bibfield  {journal} {\bibinfo  {journal}
  {Nature materials}\ }\textbf {\bibinfo {volume} {12}},\ \bibinfo {pages}
  {507} (\bibinfo {year} {2013})}\BibitemShut {NoStop}%
\bibitem [{\citenamefont {Chaudhuri}\ \emph {et~al.}(2007)\citenamefont
  {Chaudhuri}, \citenamefont {Berthier},\ and\ \citenamefont
  {Kob}}]{Kob2007universal}%
  \BibitemOpen
  \bibfield  {author} {\bibinfo {author} {\bibfnamefont {P.}~\bibnamefont
  {Chaudhuri}}, \bibinfo {author} {\bibfnamefont {L.}~\bibnamefont {Berthier}},
  \ and\ \bibinfo {author} {\bibfnamefont {W.}~\bibnamefont {Kob}},\
  }\href@noop {} {\bibfield  {journal} {\bibinfo  {journal} {Physical review
  letters}\ }\textbf {\bibinfo {volume} {99}},\ \bibinfo {pages} {060604}
  (\bibinfo {year} {2007})}\BibitemShut {NoStop}%
\bibitem [{\citenamefont {Weeks}\ \emph {et~al.}(2000)\citenamefont {Weeks},
  \citenamefont {Crocker}, \citenamefont {Levitt}, \citenamefont {Schofield},\
  and\ \citenamefont {Weitz}}]{weeks2000three}%
  \BibitemOpen
  \bibfield  {author} {\bibinfo {author} {\bibfnamefont {E.~R.}\ \bibnamefont
  {Weeks}}, \bibinfo {author} {\bibfnamefont {J.~C.}\ \bibnamefont {Crocker}},
  \bibinfo {author} {\bibfnamefont {A.~C.}\ \bibnamefont {Levitt}}, \bibinfo
  {author} {\bibfnamefont {A.}~\bibnamefont {Schofield}}, \ and\ \bibinfo
  {author} {\bibfnamefont {D.~A.}\ \bibnamefont {Weitz}},\ }\href@noop {}
  {\bibfield  {journal} {\bibinfo  {journal} {Science}\ }\textbf {\bibinfo
  {volume} {287}},\ \bibinfo {pages} {627} (\bibinfo {year}
  {2000})}\BibitemShut {NoStop}%
\bibitem [{\citenamefont {Bak}\ \emph {et~al.}(1988)\citenamefont {Bak},
  \citenamefont {Tang},\ and\ \citenamefont {Wiesenfeld}}]{bak1988self}%
  \BibitemOpen
  \bibfield  {author} {\bibinfo {author} {\bibfnamefont {P.}~\bibnamefont
  {Bak}}, \bibinfo {author} {\bibfnamefont {C.}~\bibnamefont {Tang}}, \ and\
  \bibinfo {author} {\bibfnamefont {K.}~\bibnamefont {Wiesenfeld}},\
  }\href@noop {} {\bibfield  {journal} {\bibinfo  {journal} {Physical review
  A}\ }\textbf {\bibinfo {volume} {38}},\ \bibinfo {pages} {364} (\bibinfo
  {year} {1988})}\BibitemShut {NoStop}%
\bibitem [{\citenamefont {Turcotte}(1999)}]{turcotte1999self}%
  \BibitemOpen
  \bibfield  {author} {\bibinfo {author} {\bibfnamefont {D.~L.}\ \bibnamefont
  {Turcotte}},\ }\href@noop {} {\bibfield  {journal} {\bibinfo  {journal}
  {Reports on progress in physics}\ }\textbf {\bibinfo {volume} {62}},\
  \bibinfo {pages} {1377} (\bibinfo {year} {1999})}\BibitemShut {NoStop}%
\bibitem [{\citenamefont {Sethna}\ \emph {et~al.}(2001)\citenamefont {Sethna},
  \citenamefont {Dahmen},\ and\ \citenamefont {Myers}}]{sethna2001crackling}%
  \BibitemOpen
  \bibfield  {author} {\bibinfo {author} {\bibfnamefont {J.~P.}\ \bibnamefont
  {Sethna}}, \bibinfo {author} {\bibfnamefont {K.~A.}\ \bibnamefont {Dahmen}},
  \ and\ \bibinfo {author} {\bibfnamefont {C.~R.}\ \bibnamefont {Myers}},\
  }\href@noop {} {\bibfield  {journal} {\bibinfo  {journal} {Nature}\ }\textbf
  {\bibinfo {volume} {410}},\ \bibinfo {pages} {242} (\bibinfo {year}
  {2001})}\BibitemShut {NoStop}%
\end{thebibliography}%

\end{document}